\documentstyle{article}
\title{\bf Misner-brane cosmology}
\author{ Pedro F. Gonz\'{a}lez-D\'{\i}az.\\
Isaac Newton Institute, 20 Clarkson Road, Cambridge, CB3 0EH,
UK\\ and\\ Instituto de Matem\'{a}ticas y F\'{\i}sica Fundamental\\
Consejo Superior de Investigaciones Cient\'{\i}ficas\\ Serrano 121,
28006 Madrid, SPAIN\\ }
\date{June 1, 2000}
\begin{document}
\maketitle
\large
\setlength{\baselineskip}{0.5cm}

\begin{center}
{\bf Abstract}
\end{center}

A brane universe derived from the Randall-Sundrum models is
considered in which an additional Misner-like periodicity is
introduced in the extra direction. This model solves the
ambiguity in the choice of the brane world by identifying the
branes with opposite tensions, in such a way that if one enters
the brane with positive tension, one finds oneself emerging from
the brane with negative tension, without having experienced any
tension. We show that the cosmological evolution resulting from
this model matches that of the standard Friedmann scenario, at
least in the radiation dominated era, and that there exist
closed timelike curves only in the bulk, but not in the branes
which are chronologically protected from causality violations by
quantum-mechanically stable chronology horizons.

\pagebreak

In his public lecture {\it Space and time warps} Stephen Hawking
has recently pointed out [1] that string theory might now be
offering a new room for possible violation of the chronology
protection conjecture [2], coming out from some special mixing
between our four flat directions of space and time and the extra
highly curved or warped directions considered in string
theories, so traveling superluminarly or back in time cannot be
ruled out yet. The present letter describes a string-theory
inspired cosmological scenario where the above issue can be
addressed properly. More precisely, we shall consider a brane
world derived from the Randall-Sundrum models (originally
intended to solve the hierarchy problem [3,4]) by introducing a
periodicity on the extra direction which solves any ambiguity in
the choice of the brane world and induces the emergence of
nonchronal regions.

Let us start with a five-dimensional spacetime with the fifth
dimension, $\omega$, compactified on S$^{1}$, with
$-\omega_{c}\leq\omega\leq\omega_{c}$, and satisfying the
orbifold symmetry $\omega\leftrightarrow-\omega$. On the fifth
direction there are two domain walls, with the brane at
$\omega=0$ having positive tension and that at $\omega=\omega_c$
having negative tension. A first model assumed [3] that we live
in the negative-tension brane where the mass scales are severely
suppressed, but had the serious problem of a repulsive gravity
[5]. A second model assumed [4] that we are living in the
positive-tension brane, while the other brane was moved off to
infinity. This avoids any repulsive character of gravity, but
unfortunately leads to field equations on the brane at
$\omega=0$ which are nonlinear in the source terms [6]. In spite
of the several attempts made to reconcile this non coventional
behaviour with the standard Friedmann scenario based on
inserting a cosmological constant in the brane universe [7-10],
within the spirit of the Randall-Sundrum approach [3,4], it
appears clear that such a situation is rather unconfortable in a
number of respects, not less of which is the fact that in any
classical gravitational theory where isotropy and homogeneity
are assumed, one should expect standard cosmology to hold.

In order to represent the universe we live in, our approach
chooses neither of the two branes on $\omega$ individually, but
both of them simultaneously; that is to say, we shall provide
the fifth direction with a periodic character, in such a way
that the branes at $\omega=0$ and $\omega=\omega_c$ are
identified with each other, so if one enters the brane at
$\omega=0$, one finds oneself emerging from the brane at
$\omega=\omega_c$, without having experienced any tension. If we
then set the brane at $\omega=0$ into motion toward the brane at
$\omega=\omega_c$ with a given speed $v$, in units of the speed
of light, our space would resemble five-dimensional Misner
space, the differences being in the spatial topology and in the
definition of time and the closed-up extra direction which would
also contract at a rate $v$. Then, time dilation between the two
branes would inexorably lead to the creation of a nonchronal
region which will start forming at the future of a given
chronology horizon. We shall first consider the metric of the
five-dimensional spacetime in terms of Gaussian coordinates
centered e.g. on the brane at $\omega=0$. If we assume the three
spatial sections on the branes to be flat, then such a metric
can be written in the form [11]
\begin{equation}
ds^2= c^2(\omega,t)\left(d\omega^2-dt^2\right)+
a^2(\omega,t)\sum_{j=2}^{4}dx_j^2 ,
\end{equation}
where if we impose the orbifold condition
$\omega\leftrightarrow-\omega$, the scale factors $c$ and $a$
are given by
\begin{equation}
c^2(\omega,t)
=\frac{\dot{f}(u)\dot{g}(v)}{\left[f(u)+g(v)\right]^{\frac{2}{3}}}
,\;\;\; a^2(\omega,t) =\left[f(u)+g(v)\right]^{\frac{2}{3}} ,
\end{equation}
with $u=t-|\omega|$ and $v=t+|\omega|$ the retarded and advanced
coordinates satisfying the orbifold symmetry, where we have
absorbed some length constants into the definition of $t$ and
$\omega$, and the overhead dot denotes derivative with respect
to time $t$. If no further symmetries are introduced then $f(u)$
and $g(v)$ are arbitrary functions of $u$ and $v$, respectively
[11] . However, taking metric (1) to also satisfy the (Misner)
symmetry [12,13]
\[\left(t,\omega,x_2,x_3,x_4\right)\leftrightarrow\]
\[\left(t\cosh(n\omega_c)+\omega\sinh(n\omega_c),
t\sinh(n\omega_c)+\omega\cosh(n\omega_c),\right.\]
\begin{equation}
\left. x_2,x_3,x_4\right)
,
\end{equation}
where $n$ is any integer number, makes the functions $f(u)$ and
$g(v)$ no longer arbitrary. Invariance of metric (1) under
symmetry (3) can be achieved if we choose for $f(u)$ and $g(v)$
e.g. the simple expressions
\begin{equation}
f(u)=\ln u ,\;\;\; g(v)=\ln v .
\end{equation}
Imposing symmetry (3) together with the choice for the scale
factors given by expressions (4) fixes the topology of the
five-manifold to correspond to the identification of the domain
walls at $\omega=0$ and at $\omega=\omega_c$ with each other, so
that if one enters one of these branes then one finds oneself
emerging from the other. Although other possible, perhaps more
complicate choices for the functions $f(u)$ and $g(v)$ could
also be done in order to achieve fulfillment of the above
symmetry, in what follows we shall restrict ourselves to use
Eqs. (4) to define our brane universe model satisfying the
Misner like symmetry (3), and denote it as the Misner-brane
universe. We shall show that the choice (4) actually corresponds
to an early universe which is radiation dominated.

The periodicity property on the extra direction can best be
explicited by introducing the coordinate transformation
\begin{equation}
\omega=T\sinh(W) ,\;\;\; t=T\cosh(W) ,
\end{equation}
with which metric (1) becomes
\begin{equation}
ds^2=
\frac{\left(\frac{\dot{T}^2}{T^2}-
\dot{W}^2\right)}{\ln^{\frac{2}{3}}T^2}\left(T^2
dW^2-dT^2\right)+ \ln^{\frac{2}{3}}T^2\sum_{j=2}^4 dx_j^2 .
\end{equation}
Although now metric (6) and the new coordinate
$T=\sqrt{t^2-\omega^2}$ (which is timelike, provided that $\ln
T\geq {\rm Const}\pm W$) are both invariant under symmetry (3),
the new extra coordinate $W$ transforms as
\begin{equation}
W\equiv\frac{1}{2}\ln\left(\frac{t+|\omega|}{t-|\omega|}\right)
\leftrightarrow W+n\omega_c
\end{equation}
under that symmetry. On the two identified branes making up the
Misner-brane universe, we can describe the four-dimensional
spacetime by a metric which can be obtained by slicing the
five-dimensional spacetime given by metric (6), along surfaces
of constant $W$, i.e.
\begin{equation}
ds^2= -\frac{\dot{T}^2}{T^2\ln^{\frac{2}{3}}T^2}dT^2+
\ln^{\frac{2}{3}}T^2\sum_{j=2}^4 dx_j^2 .
\end{equation}
The energy-momentum tensor of this brane universe will now have
the form:
\begin{equation}
T_i^k= \frac{\delta(\omega-n\omega_c)}{c_b}{\rm
diag}\left(-\rho,p,p,p,0\right),\;\; n=0,1,2,3... ,
\end{equation}
where $c_b\equiv(t,\omega=n\omega_c)$. This tensor should be
derived using the Israel's jump conditions [14] that follow from
the Einstein equations. Using the conditions computed by
Bin\'{e}truy, Deffayet and Langlois [6] and the metric (6) we then
[11] obtain for the energy density and pressure of our
Misner-brane universe:
\begin{equation}
\rho= -\frac{4T\dot{W}}{\kappa_{(5)}^2\ln^{\frac{2}{3}}T^2
\left(|\dot{T}^2-T^2\dot{W}^2|\right)^{\frac{1}{2}}}
\end{equation}
\begin{equation}
p=\frac{2T\dot{T}^2\ln^{\frac{2}{3}}T^2}{\kappa_{(5)}^2
\left(|\dot{T}^2-T^2\dot{W}^2|\right)^{\frac{5}{2}}}
\frac{d}{dt}\left(\frac{T\dot{W}}{\dot{T}}\right)-\frac{1}{3}\rho .
\end{equation}
Thus, both the energy density $\rho$ and the pressure $p$,
defined by expressions (10) and (11), respectively, identically
vanish on the sections $W$=const. Therefore, taking the jump of
the component ($\omega,\omega$) of the Einstein equations with
the orbifold symmetry [6], one gets on the identified branes
\begin{equation}
\frac{\dot{a}_b^2}{a_b^2} +\frac{\ddot{a}_b}{a_b}=
\frac{\dot{a}_b\dot{c}_b}{a_b c_b} ,
\end{equation}
where $a_b\equiv a(t,\omega=n\omega_c)$, with $n=0,1,2,3...$, is
the scale factor in our Misner-brane universe.

The breakdown of arbitrariness of functions $f(u)$ and $g(v)$
imposed by symmetry (3) prevents the quantity $c_b$ to be a
constant normalizable to unity, so the right-hand-side of Eq.
(12) can be expressed in terms of coordinates $T,W$ as:
\begin{equation}
\frac{\dot{a}_b\dot{c}_b}{a_b c_b}= -\frac{1+\frac{1}{3\ln
T}}{3T^2\cosh^2 W\ln T} .
\end{equation}
A simple dimensional analysis (performed after restoring the
constants absorbed in the definitions of $t$ and $\omega$ in
Eqs. (2)) on the right-hand-side of Eq. (13) indicates that if
this side is taken to play the role of the source term of the
corresponding Friedmann equation, then it must be either
quadratic in the energy density if we use
$\kappa_{(5)}^2=M_{(5)}^{-3}$ (with $M_{(5)}$ the
five-dimensional reduced Planck mass) as the gravitational
coupling, or linear in the energy density and pressure if we use
$\kappa_{(4)}^2=8\pi G_N=M_{(4)}^{-2}$ (with $M_{(4)}$ the usual
four-dimensional reduced Planck mass) as the gravitational
coupling. Since $\kappa_{(4)}^2$ should be the gravitational
coupling that enters the (Friedmann-) description of our
observable four-dimensional universe, we must choose the
quantity in the right-hand-side of Eq. (13) to represent the
combination $-\kappa_{(4)}^2(\rho_b+3p_b)/6$ which should be
associated with the geometrical left-hand-side part of Eq. (12)
of the corresponding Friedmann equation, when the term
proportional to the bulk energy-momentum tensor
$T_{\omega\omega}$ is dropped by taking the bulk to be empty
(see later on). We have then,
\begin{equation}
\rho_b+3p_b =\frac{2\left(1+\frac{1}{3\ln
T}\right)}{\kappa_{(4)}^2 T^2\cosh^2 W\ln T} .
\end{equation}
The four-dimensional metric (8) can be expressed as that of a
homogeneous and isotropic universe with flat spatial geometry,
$ds^2=-d\eta^2+a(\eta)_b^2\sum_{j=2}^4 dx_j^2$, if we take for
the cosmological time $\eta=3a(\eta)_b^2/(4\cosh
W)=3\ln^{2/3}T^2/(4\cosh W)$. In this case, the scale factor
$a(\eta)_b$ corresponds to that of a radiation dominated flat
universe, with $\cosh W=const$ expressing conservation of rest
energy, and $p_b=\rho_b/3$ at small $\eta$. For small $\eta$, it
follows then from Eq. (14)
\[\rho_b\equiv \rho_b(T,\eta)\simeq\frac{4}{3\kappa_{(4)}^2
T^2\cosh^2(W)a(\eta)_b^6} ,\] or
\[\rho_b(\eta)=a(\eta)_b^2 T^2\cosh^2 W\rho_b(T,\eta)\simeq
\frac{3}{32\pi G_N\eta^2} ,\]
when expressed in terms of the cosmological time $\eta$ only.

Having thus shown that the Misner-brane cosmology based on
ansatz (4) matches the standard cosmological evolution in the
radiation dominated era, we turn now to investigate the
nonchronal character of the spacetimes described by metric (6).
Nonchronal regions in such spacetimes can most easily be
uncovered if we re-define the coordinates entering this metric,
such that $Y=W-\ln T$ and $\Theta=T^2$. In terms of the new
coordinates, the line element (6) reads:
\begin{equation}
ds^2= -\frac{\left(\dot{Y}^2
+\frac{\dot{Y}\dot{\Theta}}{\Theta}\right)}{\ln^{\frac{2}{3}}\Theta}
\left(\Theta dY^2
+dYd\Theta\right)+\ln^{\frac{2}{3}}\Theta\sum_{j=2}^4 dx_j^2 .
\end{equation}
This metric is real only for $\Theta>0$ in which case $Y$ is
always timelike if $\dot{Y}>0$. One will therefore [13] have
closed timelike curves (CTC's) only in the bulk, provided
$\Theta>0$, $\dot{Y}>0$. There will never be CTC's in any of the
branes, that is the observable universe.

Singularities of metrics (6), (8) and (15) will appear at $T=0$
and $T=1$. The first one corresponds to $\omega=t=0$, and the
second one to $\eta=0$, the initial singularity at $Y=W$,
$t^2=1+\omega^2$, in a radiation dominated universe. We note
that the source term $-\kappa_{(4)}^2(\rho_b+3p_b)/6$ given by
Eq. (14) also diverges at these singularities. The geodesic
incompleteness at $T=1$ can be removed in the five-dimensional
space, by extending metric (6) with coordinates defined e.g by
$X=\int dW/\ln^{\frac{1}{3}}T^2-3\ln^{\frac{2}{3}}T^2/4$,
$Z=\int dW/\ln^{\frac{1}{3}}T^2+3\ln^{\frac{2}{3}}T^2/4$.
Instead of metric (6), we obtain then
\begin{equation}
ds^2=
\frac{2}{3}(Z-X)
\left\{\exp\left[\sqrt{\frac{8}{27}}
(Z-X)^{\frac{3}{2}}\right]\dot{X}\dot{Z}dXdZ +\sum_{j=2}^4
dx_j^2\right\} ,
\end{equation}
where one can check that whereas the singularity at $T=0$ still
remains, the metric is now regular at $T=1$. Since replacing $W$
for $Y$ in Eqs. (6) and (15) simultaneously leads to the
condition $Y=-\frac{1}{2}\ln T +const.$, and hence, by the
definition of $Y$, $Y=const$ and $W=const$ at $T=1$, one can
choose the singularities at $T=1$ to correspond to the brane
positions along $\omega$, and interpret such singularities as
chronology horizons in the five-space. So, CTC's will only
appear in the bulk on nonchronal regions defined by $0<T<1$,
$n\omega_c
<W<(n+1)\omega_c$, with $n=0,1,2,3,...$. The resulting scenario
can be regarded to be a typical example of the kind of models
alluded by Hawking [1] on how curved or warped extra dimensions
would induce the existence of nonchronal regions in higher
dimensional cosmological spacetimes inspired in string theory.
In the present case, the extra direction is mixed up with our
four dimensions in such a way that, although CTC's are allowed
to occur in the bulk, any violation of the chronology protection
conjecture is fully prevented in the observable universe by the
big-bang singularity itself.

For any equation of state the combination $\rho_b+3p_b$ given in
Eq. (14) is divergent at the geodesic incompletenesses at $T=0$
and $T=1$. The classical divergence at the chronology horizons
is of course removed in the extended coordinate frame $X,Z$.
However, as it happens in wormholes [15] and other topological
generalizations [16-18] of the Misner space, if one considers a
quantum field propagating in our spacetime, then the
renormalized stress-energy tensor $\langle
T_{\mu\nu}\rangle_{ren}$ would diverge at the chronology
horizons [19]. The existence of this semiclassical instability
would support a chronology protection conjecture also against
the existence of our universe model. Two situations have been
however considered where that conjecture is violated. Both of
them use an Euclidean continuation and lead to a vanishing
renormalized stress-energy tensor everywhere, even on the
chronology horizons. In what follows, we briefly review them, as
adapted to our present problem. In order to convert metric (16)
into a positive definite metric, it is covenient to use new
coordinates $p, q$, defined by $X=p-q$, $Z=p+q$, or
$T^2=\exp\left[(4q/3)^{3/2}\right]$, $W^2=4p^2 q/3$. A positive
definite metric is then obtained by the continuation $p=i\xi$
which, in turn, implies $W=i\Omega$. Furthermore, using Eqs. (5)
we can also see that this rotation converts the extra direction
$\omega$ in pure imaginary and keeps $t$ and $T$ real, while
making the first two of these three quantities periodic and
leaving $T$ unchanged. Two ans\"atze can then be used to fix the
value of $P_{\Omega}$, the period of $\Omega$ in the Euclidean
sector. On the one hand, from $\exp(W)\rightarrow\exp(i\Omega)$
we obtain $P_{\Omega}=2\pi$, a result that allows us to
introduce a self-consistent Li-Gott vacuum [20], and hence
obtain $\langle T_{\mu\nu}\rangle_{ren}=0$ everywhere. On the
other hand, if we take $\exp(p)\rightarrow\exp(i\xi)$, then we
get $P_{\Omega}=2\pi\ln^{1/3}T^2$. In this case, for an
automorphic scalar field $\phi(\gamma X,\alpha)$, where $\gamma$
represents symmetry (3), $\alpha$ is the automorphic parameter,
$0<\alpha <1/2$, and $X=t,\omega,x^2,x^3,x^4$, following the
analysis carried out in [21,22], one can derive solutions of the
field equation $\Box\phi=\Box\bar{\phi}=0$ by demanding
$t$-independence for the mode-frequency. This amounts [22] to a
quantum condition on time $T$ which, in this case, reads $\ln
T^2=(n+\alpha)^3\ln T_0^2$, where $T_0$ is a small constant
time. The use of this condition in the Hadamard function leads
to a value for $\langle T_{\mu\nu}\rangle_{ren}$ which is again
vanishing everywhere [22]. This not only solves the problem of
the semiclassical instability, but can also regularize
expression (16) at $T=0$ and $T=1$:
\begin{equation}
\rho_b+2p_b=
\frac{2T_0^{-2(n+\alpha)^3}\left(1+\frac{1}{3(n+\alpha)^3\ln
T_0}\right)}{\kappa_{(4)}^2\cosh^2(W) (n+\alpha)^3\ln T_0} ,
\end{equation}
which can never diverge if we choose the constant $T_0$ such
that $\ln T_0\neq 0$.

At first sight, it could seem that Misner symmetry describes
simple and familiar spacetimes. Specifically one would believe
this by showing that Misner symmetry converts the
five-dimensional metric (1) into merely a reparametrization of
the Kasner-type solution [23]. However, the simple
transformation $Q=\ln T$ converts metric (1) into
\[ds^2=(2Q)^{-2/3}\dot{u}\dot{v}\left(-dQ^2+dW^2\right)
+(2Q)^{2/3}\sum_{j=2}^{4}dx_j^2 ,\] which differs from a
Kasner-type metric by the factor
$\dot{u}\dot{v}=1-|\dot{\omega}|^2$ in the first term of the
right-and-side. This factor cannot generally be unity in the
five-dimensional manifold. On surfaces of constant $W=W_0$,
according to Eqs. (5), we have $\dot{\omega}=\tanh W_0$, so
$\dot{u}\dot{v}=\cosh^{-2}W_0$ which can only be unity for
$W_0=0$ that is on the brane at $\omega=0$. However, besides
identifying the two branes according to Eq. (7), the Misner
approach also requires that the closed up direction $\omega$
contracts at a given nonzero rate $d\omega_c/d\eta= -v_0$ [24].
This in turn means that once the branes are set in motion toward
one another at the rate $v_0$, symmetry (7) should imply that
for constant $W_0$,
\[\frac{dW_0(0)}{d\eta}=0\leftrightarrow\frac{dW_0(\eta)}{d\eta}
-nv_0=0 ,\] so that $dW_0(\eta)/d\eta\neq 0$ if $n\neq 0$. In
this case, we have $\bigtriangleup
W_0(\eta)=n\int_0^{\eta}d\omega_c=n\bigtriangleup_{\eta}\omega_c$,
and hence $W_0(\eta)=W_0(0)+n\bigtriangleup_{\eta}\omega_c=
n\bigtriangleup_{\eta}\omega_c >0$, provided that we initially
set $W_0\equiv W_0(0)=0$. It follows that $\dot{u}\dot{v}$ can
only be unity on the brane at $\omega=0$ when $n=0$ (i.e. at the
very moment when the brane universe was created and started to
evolve. We note that if we substract the zero-point contribution
$\alpha\ln^{1/3}T_0^2$, the quantization of $T$ discussed above
amounts to the relation $\eta\propto n^2$ and, therefore,
initial moment at $\eta=0$ means $n=0$), taking on
smaller-than-unity values thereafter, to finally vanish as
$\eta,n\rightarrow\infty$. Thus, one cannot generally consider
metric (1) or metrics (6) and (8) to be reparametrizations of
the Kasner solution neither in five nor in four dimensions,
except at the very moment when brane at $\omega=0$ starts being
filled with radiation, but not later even on this brane.

We note that in the case that Kasner metric would exactly
describe our spacetime (as it actually happens at the classical
time origin, $T=1$, $n=0$), Misner identification reduces to
simply identifying the plane $W=0$ with $W=n\omega_c$, that is
identifying $W$ on a constant circle, which does not include
CTC's. This picture dramatically changes nevertheless once $n$
and $\eta$ become no longer zero, so that
$\dot{u}\dot{v}=\cosh^{-2}W_0
< 1$ and the metric cannot be expressed as a reparametrization
of the Kasner metric. In that case, there would appear a past
apparent singularity [actually, a past event (chronology)
horizon] at $T=1$ for observers at later times $\eta,n\neq 0$,
which is extendible to encompass nonchronal regions containing
CTC's, as showed before by using the extended metric (16).
Indeed, the particular value of $T$-coordinate $T=1$ measures a
quantum transition at which physical domain walls (three-branes)
with energy density $\rho_b$ created themselves, through a
process which can be simply represented by the conversion of the
inextendible physical singularity of Kasner metric [23] at
$T=1$, $n=0$ into the coordinate singularity of the Misner-brane
metric at $T=1$, relative to observers placed at later times
$\eta,n\neq 0$, which is continuable into a nonchronal region on
the bulk space.

On the other hand, since the energy density $\rho$ and pressure
$p$ on any of the two candidate branes vanish, one might also
think that, related to the previous point, we are actually
dealing with a world with no branes, but made up enterely of
empty space. The conversion of the field-equation term (13) in a
stress-energy tensor would then simply imply violation of
momentum-energy conservation. However, the existence of an event
(chronology) horizon which is classically placed at $T=1$ for
the five-dimensional spacetime amounts to a process of quantum
thermal radiation from vacuum, similar to those happening in
black holes or de Sitter space [25,26], which observers at later
times $\eta, n>0$ on the branes would detect to occur at a
temperature $\beta\propto\ln^{-1/3}T^2$, when we choose for the
period of $\Omega$ (which corresponds to the Euclidean
continuation of the {\it timelike} coordinate $W$ on
hypersurfaces of constant $T$)
$P_{\Omega}=2\pi\ln^{1/3}T^2\propto a$. Thus, for such
observers, the branes would be filled with radiation having an
energy density proportional to
$\ln^{-4/3}T^2\propto\eta^{-2}=\rho_b$ and temperature
$\propto\eta^{-1/2}$, i.e. just what one should expect for a
radiation dominated universe and we have in fact obtained from
Eq. (14). Observers on the branes at times corresponding to
$T>1$, $n\neq 0$ would thus interpret all the radiating energy
in the four-dimensional Misner-brane universe to come from
quantum-mechanical particle creation near an event horizon at
$T=1$.

Moreover, in order to keep the whole two-brane system
tensionless relative to a {\it hypothetical} observer who is
able to pass through it by tunneling along the fifth dimension
(so that when the observer enters the brane at $\omega=0$ she
finds herself emerging from the brane at $\omega=\omega_c$,
without having experienced any tension), one {\it must} take the
tension $V_{\omega=0}=\rho_b>0$ and the tension
$V_{\omega=\omega_c}=-\rho_b$, and therefore the total tension
experienced by the hypothetical observer,
$V=V_{\omega=0}+V_{\omega=\omega_c}$ will vanish. Given the form
of the energy density $\rho_b$, this necessarily implies that
current observers should live on just one of the branes (e.g. at
$\omega=0$) and cannot travel through the fifth direction to get
in the other brane (so current observers are subjected to
chronology protection [2]), and that, relative to the
hypothetical observer who is able to make that traveling, the
brane which she emerges from (e.g. at $\omega=\omega_c$) must
then be endowed with an antigravity regime with $G_N <0$ [5],
provided she first entered the brane with $G_N >0$ (e.g. at
$\omega=0$).

Chronology protection conjecture states [2] that the laws of
physics prevent the existence of CTC's and possible time
machines constructed out of them, at least in a semiclassical
approximation where the quantum fields propagate in a classical
background spacetime. As formulated in this way, this conjecture
is violated in our model and, indeed, in all nonchronal
spacetime models admitting similar Euclidean continuations
[20-22]. The results of the present letter imply, nevertheless,
that, although the laws of physics actually allow CTC's and time
machines to occur, they place them outside our observable
universe, in such a way that such constructs can neither be
directly observed, nor break causality. It is in this sense that
a chronology protection must be understood in the present work.

To sum up, we have considered a brane universe in which the two
domain walls on the fifth direction of the Randall-Sundrum
approach are identified by using a Misner-like symmetry,
resulting in a cosmological evolution which matches that of the
standard Friedmann scenario at early times. This universe model
has CTC's in the bulk, but not in the branes which are
chronologically protected by quantum-mechanically stable
chronology horizons, so providing the chronology protection
conjecture with a new, less demanding interpretation.

\vspace{.8cm}

\noindent {\bf Acknowledgements}

\noindent For helpful comments and a careful reading of the manuscript,
the author thanks C. Sig\"uenza. This research was supported by
DGICYT under Research Projects Nos. PB97-1218 and PB98-0520.

\noindent\section*{References}
\begin{description}

\item [1] S.W. Hawking, {\it Space and Time Warps}, public
lecture,\\ http://www.hawking.org.uk/lectures/lindex.html
\item [2] S.W. Hawking, Phys. Rev. D46(1992)603.
\item [3] L. Randall and R. Sundrum, Phys. Rev. Lett. 83(1999)3370.
\item [4] L. Randall and R. Sundrum, Phys. Rev. Lett. 83(1999)4690.
\item [5] T. Shiromizu, K. Maeda and M. Sasaki, {\it Einstein
Equations on the 3-Brane World}, gr-qc/9910076.
\item [6] P. Bin\'{e}truy, C. Deffayet and D. Langlois, Nucl. Phys.
B565(2000)269.
\item [7] C. Cs\'{a}ki, M. Graesser, C. Kolda and J. Terning, Phys.
Lett. B462(1999)34.
\item [8] J.M. Cline, C. Grosjean and G. Servant, Phys. Rev.
Lett. 83(1999)4245.
\item [9] P. Bin\'{e}truy, C. Deffayet, U. Ellwanger and D.
Langlois, Phys. Lett. B477(2000)285.
\item [10] P.F. Gonz\'{a}lez-D\'{\i}az, Phys. Lett. B481(2000)353.
\item [11] D.N. Vollick, {\it Cosmology on a three-brane},
hep-th/9911181.
\item [12] C.W. Misner, in {\it Relativity and Cosmology},
edited by J. Ehlers (American Mathematical Society, Providence,
RI, USA, 1967).
\item [13] S.W. Hawking and G.F.R. Ellis, {\it The large
scale structure of space-time} (Cambridge University Press,
Cambridge, UK, 1973).
\item [14] W. Israel, Nuovo Cimento B44(1966)1; B48(1966)463.
\item [15] M.S. Morris, K.S. Thorne and U. Yortsever, Phys.
Rev. Lett. 61(1988)1446.
\item [16] J.R. Gott, Phys. Rev. Lett. 66(1991)1126.
\item [17] P.F. Gonz\'{a}lez-D\'{\i}az, Phys. Rev. D54(1996)6122.
\item [18] P.F. Gonz\'{a}lez-D\'{\i}az and L.J. Garay, Phys. Rev. D59(1999)064026.
\item [19] S.-W. Kim and K.S. Thorne, Phys. Rev. D43(1991)3929.
\item [20] Li-Xin-li and J.R. Gott, Phys. Rev. Lett. 80(1998)2980.
\item [21] S.V. Sushkov, Class. Quantum Grav. 14(1997)523.
\item [22] P.F. Gonz\'{a}lez-D\'{\i}az, Phys. Rev. D58(1998)124011.
\item [23] R.M. Wald, {\it General Relativity} (The University of
Chicago Press, Chicago, USA, 1984).
\item [24] K.S. Thorne, in {\it Directions in General
Relativity}, Proceedings of the 1993 International Symposium,
Maryland, in Honor of C. Misner (Cambridge University Press,
Cambridge, UK, 1993), Vol. 1.
\item [25] S.W. Hawking, Commun. Math. Phys. 43(1975)199.
\item [26] G.W. Gibbons and S.W. Hawking, Phys. Rev.
D15(1977)2738.

\end{description}

\end{document}